\documentclass[traditabstract]{aa}  
\usepackage{natbib}
\usepackage{graphicx}
\usepackage{txfonts}
\usepackage{subfig}
\usepackage{hyperref}  
\usepackage{xcolor}
\usepackage{orcidlink}
\usepackage{booktabs}
\usepackage{csquotes}
\usepackage{soul}

\newcommand{\HI}{$\ion{H}{i}$}

\newcommand{\Ha}{$\mathrm{H\alpha}$}
\newcommand{\Hb}{$\mathrm{H\beta}$}

\begin{document}

   \title{ALMA-JELLY II. Constraining the radial profiles of the quenching timescales in ram-pressure-stripped galaxies}

   \subtitle{}

   \titlerunning{ALMA-JELLY II: Quenching timescales in RPS galaxies}
   \authorrunning{R.P. Infurna et al.}

  \author{Roberto P. Infurna$\,$\orcidlink{0009-0003-3608-0647} \inst{1}\thanks{\href{mailto:r.infurna1@campus.unimib.it}{r.infurna1@campus.unimib.it}},
    Matteo Fossati$\,$\orcidlink{0000-0002-9043-8764} \inst{1,2}\thanks{\href{mailto:matteo.fossati@unimib.it}{matteo.fossati@unimib.it}},
    Alessandro Boselli$\,$\orcidlink{0000-0002-9795-6433}\inst{3,4},
    Michele Fumagalli$\,$\orcidlink{0000-0002-9043-8764}\inst{1,5},
    Jeffrey D.P. Kenney$\,$\orcidlink{0000-0003-0586-6754}\inst{6},
    Pavel Jáchym$\,$\orcidlink{0000-0002-1640-5657}\inst{7},
    Ming Sun$\,$\orcidlink{0000-0001-5880-0703}\inst{8},
    Elias Brinks$\,$\orcidlink{0000-0002-7758-9699}\inst{9},
    Françoise Combes$\,$\orcidlink{0000-0003-2658-7893}\inst{10},
    Patricia da Silva$\,$\orcidlink{0000-0001-8837-8670}\inst{8},
    Romana Grossová$\,$\orcidlink{0000-0003-3471-7459}\inst{7},
    Stephen Gwyn$\,$\orcidlink{0000-0001-8221-8406}\inst{11},
    Kokoro Hosogi$\,$\orcidlink{0009-0009-7710-5760}\inst{12},
    Alessandro Ignesti$\,$\orcidlink{0000-0003-1581-0092}\inst{7},
    Jin Koda\inst{13},
    Jan Palouš\inst{7},
    Nathanael Pichette$\,$\orcidlink{0009-0002-6417-3448}\inst{8},
    Sujith Ranasinghe\inst{14},
    Tom C. Scott$\,$\orcidlink{0000-0002-3746-4859}\inst{15},
    Harrison J. Souchereau$\,$\orcidlink{0000-0001-5079-1865}\inst{6},
    Juhi Tiwari$\,$
    \orcidlink{0000-0002-7705-4483}\inst{8},
    Bernd Vollmer\inst{16},
    Norbert Werner$\,$\orcidlink{0000-0003-0392-0120}\inst{17},
    Masafumi Yagi$\,$\orcidlink{0000-0001-7550-2281}\inst{18}
}

\institute{ 
        Dipartimento di Fisica ``G. Occhialini'', Universit\`a di Milano-Bicocca, Piazza della Scienza 3, 20126 Milano, Italy
    \and
        INAF - Osservatorio Astronomico di Brera, via Brera 28, 20121 Milano, Italy
	\and    
        Aix Marseille Univ, CNRS, CNES, LAM, Marseille, France
    \and
        INAF - Osservatorio Astronomico di Cagliari, Via della Scienza 5, 09047 Cuccuru Angius, Selargius CA, Italy
    \and 
        INAF - Osservatorio Astronomico di Trieste, Via G.B. Tiepolo, 11 34143 Trieste, Italy
    \and
        Department of Astronomy, Yale University, New Haven, CT 06511, USA
    \and 
        Astronomical Institute of the Czech Academy of Sciences, Bo\v{c}n\'{i} II 1401, 141 00, Prague, Czech Republic
    \and 
        Department of Physics and Astronomy, The University of Alabama in Huntsville, 301 Sparkman Dr, Huntsville, AL 35899, USA
    \and University of Hertfordshire Centre for Astrophysics Research, College Lane, Hatfield, AL10      9AB United Kingdom
    \and Observatoire de Paris, LUX, Collège de France, CNRS, PSL University, Sorbonne University,       75014, Paris, France
    \and Canada-France-Hawaii Telescope, 65-1238 Mamalahoa Highway, Kamuela, HI 96743, USA
    \and The Leonard E. Parker Center for Gravitation, Cosmology and Astrophysics, Department of Physics \& Astronomy, University of Wisconsin-Milwaukee, 3135 N Maryland Avenue, Milwaukee, WI 53211, USA
    \and Stony Brook University, Stony Brook, NY 11794-3800, USA
    \and Department of Physics and Astronomy, University of Calgary, Calgary, AB T2N 1N4, Canada    
    \and Institute of Astrophysics and Space Sciences (IA), Rua das Estrelas, 4150-762 Porto, Portugal
    \and Universit\'{e} de Strasbourg, CNRS, Observatoire Astronomique de Strasbourg, UMR 7550, 67000 Strasbourg, France
    \and Department of Theoretical Physics and Astrophysics, Masaryk University, Brno, 61137, Czech Republic
    \and National Astronomical Observatory of Japan, Osawa, Mitaka, Tokyo 181-8588, Japan
    }    
                      
   \date{Received XX; accepted XX}

\abstract{
We investigate the environmental quenching histories and spatial progression of ram-pressure stripping in a sample of 11 galaxies across the nearby Coma, Norma, and A1367 clusters. As part of the ALMA-JELLY large program, we combine high-sensitivity VLT/MUSE integral-field spectroscopy with multi-wavelength photometry spanning from the ultraviolet to the far-infrared across 37 spatially resolved disc apertures. The sample spans nearly three orders of magnitude in stellar mass ($\log(M_*/\mathrm{M}_\odot) \sim 7.5 - 10.5$), with targets selected to be subject to active or recent gas removal by ram-pressure, as seen by prominent \Ha\ or UV tails. Using a Bayesian spectrophotometric fitting code, we model star formation histories featuring an exponential decline to parametrize the quenching due to ram pressure. Mock-data simulations confirm that combining integral-field spectra with broadband photometry is crucial to break degeneracies between model parameters to derive more accurate quenching ages. In all 11 galaxies, outer disc regions display halted star formation and relatively old ($\lesssim 1$ Gyr) stellar populations. For the most massive galaxies ($\log(M_*/\mathrm{M}_\odot) > 9.5$), we identify clear radial gradients in quenching ages, where inner regions quench $\approx 500~\mathrm{Myr}$ later than the outskirts, providing direct observational evidence of a prolonged outside-in stripping process regulated by their strong gravitational potential. There is surprisingly little scatter in the quenching age as a function of distance from the truncation radius, across different systems, despite variations in the cluster environment, the inclination and wind angles, and the gas distribution in the galaxy during the stripping. Conversely, in low-mass and dwarf systems ($\log(M_*/\mathrm{M}_\odot) \le 9.5$), radial age gradients nearly vanish, with disc quenching occurring more rapidly and typically within $300~\mathrm{Myr}$.}

\keywords{Galaxies: interactions, Galaxies: clusters: general, Galaxies: evolution, Techniques: imaging spectroscopy}

\maketitle
%

\section{Introduction}

It is well established that early-type galaxies are preferentially found in dense cluster environments, whereas late-type objects are more common in the field \citep{Dressler_1980}. This environmental dependence indicates that galaxy evolution is strongly influenced by local density. Several physical mechanisms have been proposed to explain the suppression of star formation in clusters, ranging from slow processes such as starvation (or strangulation) \citep{Larson_1980}, which suppress the accretion of fresh gas, to more rapid ones capable of dynamically removing the cold gas reservoir thus suppressing star formation on much shorter timescales.
In the starvation scenario, galaxies form stars for several Gyr after cosmological gas accretion has ceased, thanks to their cold gas reservoirs. 
Rapid environmental mechanisms are thought to operate in parallel with starvation, acting on timescales $\lesssim 1$ Gyr \citep{Oman_Hudson_2016} and thereby preserving the observed bimodality of the galaxy population.

Rapid quenching processes in clusters can be broadly divided into gravitational and hydrodynamical interactions. Gravitational mechanisms, such as tidal interactions and galaxy harassment, might suppress star formation by heating the gas or enhancing gas depletion \citep{Moore_1998}, although there is not much observational evidence that this is an important quenching mechanism in clusters. Hydrodynamical processes, including ram-pressure stripping (RPS) \citep{GunnGott1972} and viscous stripping \citep{Nulsen1982}, instead remove gas directly from galactic discs.

Ram pressure arises from the interaction between the hot ($10^7-10^8$ K), moderately dense ($\sim10^{-3}$ particles cm$^{-3}$) intracluster medium (ICM) and the cold interstellar medium (ISM) of a galaxy moving at high velocity through the cluster environment \citep{RPS_Review}. Since star formation requires cold gas, its removal naturally leads to quenching. Unlike gravitational perturbations, which affect both stars and gas, hydrodynamical interactions primarily affect the gaseous component while leaving the stellar distribution mostly undisturbed \citep[see e.g.][]{Consolandi_2017,Matijevic2026}. A characteristic signature of RPS is therefore the presence of a truncated gas disc with a one-sided gas tail together with a relatively undisturbed stellar disc \citep{Kenney_2004, Boselli_NGC4569_2006}. Gas truncation occurs because ram pressure becomes effective in the outer regions of galaxies at first, where the gravitational restoring force is weaker. By progressing into the inner regions, RPS thus gives rise to measurable stellar population age gradients across the galaxy discs \citep[see e.g.][]{Boselli_2016_quenching_timescales, Fossati_2018, Cramer_2019}.

The rapid suppression of star formation leaves characteristic spectroscopic signatures in the stellar population marked by strong Balmer absorption lines, more significantly H$\delta$, but also H$\beta$ and H$\alpha$, which are most often observed, being in the redder part of the optical spectrum \citep{Boselli_2016_quenching_timescales}. 
These galaxies, often referred to as Post-starburst (PSB) galaxies \citep{Dressler_1980,Poggianti_1999}, are often associated with starbursts followed by rapid quenching \citep{Crowl_2006, Kenney_2014}; however, similar signatures can also arise from abruptly truncated star formation histories without a preceding burst \citep{Boselli_2016_quenching_timescales,Vulcani_2020}. PSB galaxies are relatively rare in low-density environments but become significantly more common in clusters \citep{Poggianti_2009,gavazzi2010, Paccagnella_2016,Paccagnella_2017}, supporting an environmental origin for their rapid quenching. Using the WINGS and OmegaWINGS cluster samples, \citet{Paccagnella_2017} showed that PSBs constitute $\sim7-15$\% of cluster galaxy populations and typically experienced star formation truncation within $\lesssim1-1.5~\mathrm{Gyr}$. The SAMI Galaxy Survey revealed that H$\delta$-strong galaxies are preferentially located within $\sim0.6\,r_{200}$ and display elevated velocity dispersions, consistent with recently accreted cluster populations \citep{Oman_Hudson_2016,Owers_2019}. 

Stellar population fitting of quenched galaxies enables the reconstruction of their star formation histories (SFHs). Two main approaches are commonly adopted: parametric SFHs, which assume an analytical form for the temporal evolution of the star formation rate \citep{Cunha_2008,Carnall_2019}, and non-parametric SFHs, which reconstruct the SFH in discrete time bins without imposing a predefined shape \citep{Leja_2019,Johnson_2021}. Model SEDs are generated from the assumed SFH using stellar population synthesis (SPS) models and are then compared with observations to constrain the underlying SFH and related physical parameters.
Following pioneering papers \citep{Gavazzi_2002,Thomas_2005}, one recent example of the parametric reconstruction of SFHs in the context of environmental quenching in a complete sample including cluster galaxies is in \citet{Boselli_2016_quenching_timescales}. These authors studied the Herschel Reference Survey (HRS), a multiwavelength survey of local galaxies, including objects in the Virgo cluster. The late-type cluster galaxies were found to have been quenched rapidly, often in less than 500 Myr.  In contrast, quiescent early-type galaxies exhibit significantly older quenching ages of $\simeq 1 - 3~\mathrm{Gyr}$. These constraints, however, were derived from integrated measurements and therefore provide no spatially resolved information on how the quenching event proceeds. Spatially resolved SED fitting combining multi-band photometry with resolved spectroscopy has been adopted by \citet{Fossati_2018} in their study of the Virgo cluster galaxy NGC 4330. These authors found a radial gradient in quenching age consistent with outside-in suppression: the outer disc was quenched $\sim500~\mathrm{Myr}$ ago, whereas quenching reached the inner $\sim5~\mathrm{kpc}$ only within the last $\sim100~\mathrm{Myr}$. The innermost regions, instead, are still forming stars, fueled by the presence of HI and H$_2$ gas reservoirs, as also occurs in other Virgo galaxies \citep{Boselli_2018_vestige,Boselli_2021}.

The advent of the MUSE \citep{Bacon_2010} integral-field spectrograph has revolutionised this field, thanks to its large field of view and unparalleled sensitivity, providing high-quality spectra throughout the full extent of local galaxy discs, including the faint outer regions where the quenching event might have started. 
\citet{Vulcani_2020} analysed eight post-starburst and recently quenched cluster galaxies from the GASP survey \citep{GASP_2017}, using spatially resolved spectroscopy. Their reconstruction of the star-formation histories revealed that star formation was generally suppressed from the outskirts inward, with quenching timescales ranging from a few tens to several hundreds of Myr. \citet{Werle_2022} studied 21 PSB galaxies in $z=0.3-0.4$ clusters, finding quenching times of the order of $100-800$~Myr and a variety of radial quenching patterns, suggesting that some of those galaxies have been subject to RPS. 

In this work, we focus on the ALMA-JELLY sample (ALMA program 2021.1.01616.L, Jachym et al. in prep.), a unique multi-wavelength effort that combines ALMA cold gas observations, optical MUSE spectroscopy, and multiwavelength photometry for a sample of 28 extreme RPS galaxies, with prominent extended \Ha{} gas tails, in 3 local galaxy clusters: Coma, Norma, and Leo (A1367). First results on an individual object are given in \cite{Souchereau_2025}. From the full sample, we select 11 objects exhibiting partially (or totally) quenched stellar discs to study their environmental quenching signatures in a spatially resolved fashion across different galaxy clusters and spanning nearly 3 dex in galaxy mass.

The paper is structured as follows. In Sect.~\ref{Sec:Data}, we describe the ALMA-JELLY sample used in this work. In Sect.~\ref{Sec:Model}, we present the SED and spectral fitting methodology. In Sect.~\ref{Sec:Simulations}, we assess the performance of the fitting code using mock data. Finally, in Sect.~\ref{Sec:Results} and ~\ref{Sec:Discussion}, we present the inferred quenching parameters for our sample and discuss the results.
Through this work we will assume a flat $\Lambda$CDM cosmology with $H_0 = 67.66~\mathrm{km~s^{-1}~Mpc^{-1}}$ and $\Omega_{m,0} = 0.30966$ (Planck18; \citealt{Planck18}).

\section{Datasets and spatially resolved extractions}\label{Sec:Data}

Our sample is made of 11 galaxies drawn from three nearby massive clusters, namely Abell 1656 (Coma), Abell 3627 (Norma), and Abell 1367 (Leo cluster, A1367 hereafter), see Table~\ref{tab:galaxy_properties}. 
Consistent with other papers of this series, we assume a luminosity distance of 100 Mpc for the Coma cluster, 69.6 Mpc for Norma, and 95.8 Mpc for A1367, corresponding to angular scales of $0.463$, $0.330$, and $0.445$ kpc per arcsec, respectively. 

Galaxies were selected from the full ALMA-JELLY sample of 28 objects, according to the following criteria:
(i) Evidence of ongoing or recent RPS, from the extensive studies performed in Coma \citep{Yagi_2010,Smith_2010}, Norma \citep{Sun_2007,Sun_2010}, and A1367 \citep{Yagi_2017,Pedrini_2022}.
(ii) Availability of integral field spectroscopy with MUSE, providing spatially resolved spectra required to constrain star formation truncation timescales. 
(iii) Presence of an extended stellar disc with truncated star formation, as indicated by the absence of \Ha\ or \Hb\ emission. This criterion excludes galaxies observed with MUSE that still show extended \Ha\ emission co-spatial with the stellar disc, such as NGC 4858 in Coma \citep{Souchereau_2025} and CGCG097–079 and CGCG097–073 in A1367 \citep{Pedrini_2022}. The strong emission lines, filling the Balmer stellar absorption features, hamper a robust estimate of the time at which the quenching event has started. 
(iv) Availability of broad-band photometry covering at least one UV, two optical bands and extending, if possible, to the NIR and FIR bands. 

The properties of the galaxies in our sample are summarised in Table \ref{tab:galaxy_properties}. In order to compare galaxies with prominent gas tails with objects at a different evolutionary phase of stripping, we include the completely quenched discs of GMP 3016, GMP 4232 and RB 199. We also include NGC 3860, a partially stripped galaxy in A1367 \citep{BlueInfallingGroup}. NGC 3860 lacks a prominent \Ha\ tail, making it different from  the other sample galaxies.

\begin{table*}
\centering

\caption{Properties of the galaxies in the sample.}
\begin{tabular}{lccccccc}
\hline
Galaxy & RA, Dec (J2000) & $z$ & Type & $\log(M_*/M_\odot)$ & $V_\text{rot}$ & $R_d$ & $r/r_{200}$ \\
      &     &      &      &             & [km\,s$^{-1}$] & [kpc] &            \\

\multicolumn{8}{l}{\textbf{Coma}} \\
\hline
NGC 4848   & 194.523, 28.243 & 0.02397 & Sab & 10.32 & 164 & 3.23 & 0.49 \\
IC 4040    & 195.158, 28.058 & 0.02559 & Sdm & 9.65  & 97  & 1.90 & 0.16 \\
GMP 4232   & 194.627, 27.564 & 0.02438 & S.. & 7.39  & 17  & 0.34 & 0.49 \\
RB 199     & 194.677, 27.760 & 0.02906 & Sm  & 9.12  & 65  & 1.28 & 0.34 \\
GMP 3779   & 194.772, 27.644 & 0.01810 & Sa  & 9.74  & 105 & 2.07 & 0.35 \\
GMP 3016   & 195.004, 28.082 & 0.02599 & S.. & 7.49  & 19  & 0.37 & 0.10 \\
D100       & 195.038, 27.866 & 0.01712 & Sc & 9.26  & 73  & 1.43 & 0.10 \\
\\
\multicolumn{8}{l}{\textbf{Norma}} \\
\hline
ESO 137-001 & 243.364, -60.764 & 0.01550 & SBc? & 9.81  & 111 & 2.19 & 0.18 \\
ESO 137-002 & 243.398, -60.865 & 0.01898 & S0?  & 10.54 & 195 & 3.80 & 0.13 \\
\\
\multicolumn{8}{l}{\textbf{A1367}} \\
\hline
UGC 6697 & 175.955, 19.969 & 0.02243 & Pec & 10.13 & 142 & 2.80 & 0.38 \\
NGC 3860 & 176.205, 19.795 & 0.01872 & Sa  & 10.53 & 193 & 3.80 & 0.08 \\
\hline
\end{tabular}
\tablefoot{Mean heliocentric redshifts are taken from NED. Stellar masses are taken from \citet{RPS_Review}. Stellar velocity $V_\text{rot}$ and scale radius $R_d$ (the scale length of the exponential disc) are computed from the stellar mass using Eqs. \ref{eq:V_scaling_MW} and \ref{eq:scale_radius_scaling_MW}, assuming a fixed spin parameter $\lambda = 0.05$. The morphological type and the projected cluster-centric position are taken from \citet{RPS_Review}, except the morphological type of D100 which is taken from \cite{Cramer_2019}.}
\label{tab:galaxy_properties}
\end{table*}

\subsection{Imaging data}
The analysis is based on multi-wavelength imaging retrieved from public archives, using exclusively processed, science-ready data products. UV imaging is provided by \emph{GALEX} in the FUV and NUV bands for galaxies in the Coma and Norma clusters, and by \emph{UVIT} for Abell~1367, covering multiple NUV and FUV filters. The angular resolution is 4.2\arcsec\ and 5.3\arcsec\ in the GALEX FUV and NUV bands, respectively, while UVIT offers a substantially higher resolution of 1.4--1.7\arcsec, depending on the filter. 
For Coma and A1367 targets, optical imaging in the \emph{ugriz} bands is drawn from SDSS DR19, supplemented by $u-$band data from the CFHT UNIONS Survey \citep{Unions_2025} or public CFHT data. For Norma and for a subset of the northern targets, high-resolution optical observations are available from the \emph{Hubble Space Telescope} archive. Mid-infrared observations are taken from \emph{Spitzer}/IRAC in its four channels at 3.6, 4.5, 5.8, and 8.0\,$\mu$m, retrieved from the Spitzer Heritage Archive (SHA). Far-infrared imaging at 70, 100, and 160\,$\mu$m is provided by \emph{Herschel}/PACS using Unimap Level~2.5 products taken from the \textit{Herschel Science Archive}.

Among the datasets considered, the PSF FWHM varies by more than two orders of magnitude, from $\lesssim$0.1\arcsec\ for the HST images to $\sim11.4$\arcsec\ in the PACS 160 $\mu$m band.
We excluded near- and mid-infrared data from 2MASS and WISE because they are too shallow to enable reliable photometry in the low-surface-brightness outskirts of the galaxies or due to insufficient angular resolution. All the datasets have been astrometrically aligned to the HST or optical ground-based images, using field stars as reference points. All imaging data are converted to physical units using mission-provided calibration factors and zero points. Typical photometric calibration uncertainties are at the 1--5\% level, depending on the instrument. Foreground Milky Way extinction is corrected using $E(B-V)$ values obtained from the \citet{SF_2011_dust} dust maps via the NASA/IPAC Infrared Science Archive. Extinction values are $E(B-V)=0.0093, 0.0187, 0.1801$ mag for Coma, Abell~1367, and Norma, respectively. 

\subsection{Integral field spectroscopic data from MUSE}
Integral-field spectroscopy is provided by MUSE observations obtained from the archive and downloaded and reduced homogeneously by the ALMA--JELLY team. All data were taken in Wide Field Mode, covering a $1\times1$~arcmin$^2$ field of view with 0.2\arcsec\ pixels, and with variable seeing conditions, with an average image quality of $\approx 1$\arcsec. The spectra cover the spectral range 4800--9300\,\AA\ at a resolving power increasing from $R\sim1800$ to $R\sim3600$. Typically, multiple MUSE pointings are required to cover the full extent of the H$\alpha$ tails. We made use of all the MUSE data available for the target galaxies and we mosaicked the individual pointings in a single final datacube using point sources as astrometric references. Individual MUSE pointings have typical integration times of 30-45 min. 
A correction for foreground extinction is applied to the datacubes along the spectral dimension using the \citet{Cardelli89} law.

H$\alpha$ and H$\beta$ emission maps are extracted by subtracting a local spectral continuum from a 10\,\AA\ narrow-band (NB) centred on the redshifted line. This procedure is used exclusively to identify the star-forming and quenched regions in each galaxy. While the width of this NB is likely too narrow to encompass the full extent of the \Ha\ emission, we selected this value to avoid contamination from the nearby [NII] lines. Based on this caveat, we do not use these products for any quantitative emission-line measurement in this work. Quenched regions are characterised by the absence of H$\alpha$ emission (below 10$^{-18}$~erg/s/cm$^{2}$/arcsec$^{2}$) and, in many cases, by well-detected Balmer absorption features. The galactic disc is defined as the region with an integrated $r$-band surface brightness brighter than 25~mag arcsec$^{-2}$. For galaxies exhibiting a bright H$\alpha$ tail, if the tail is projected in front of the galactic disc (i.e., in receding galaxies), the Balmer absorption lines in the stellar spectra become contaminated by this emission. In these cases, we exclude these regions from the analysis.

\begin{figure*}
    \centering
    \includegraphics[width=1.\linewidth]{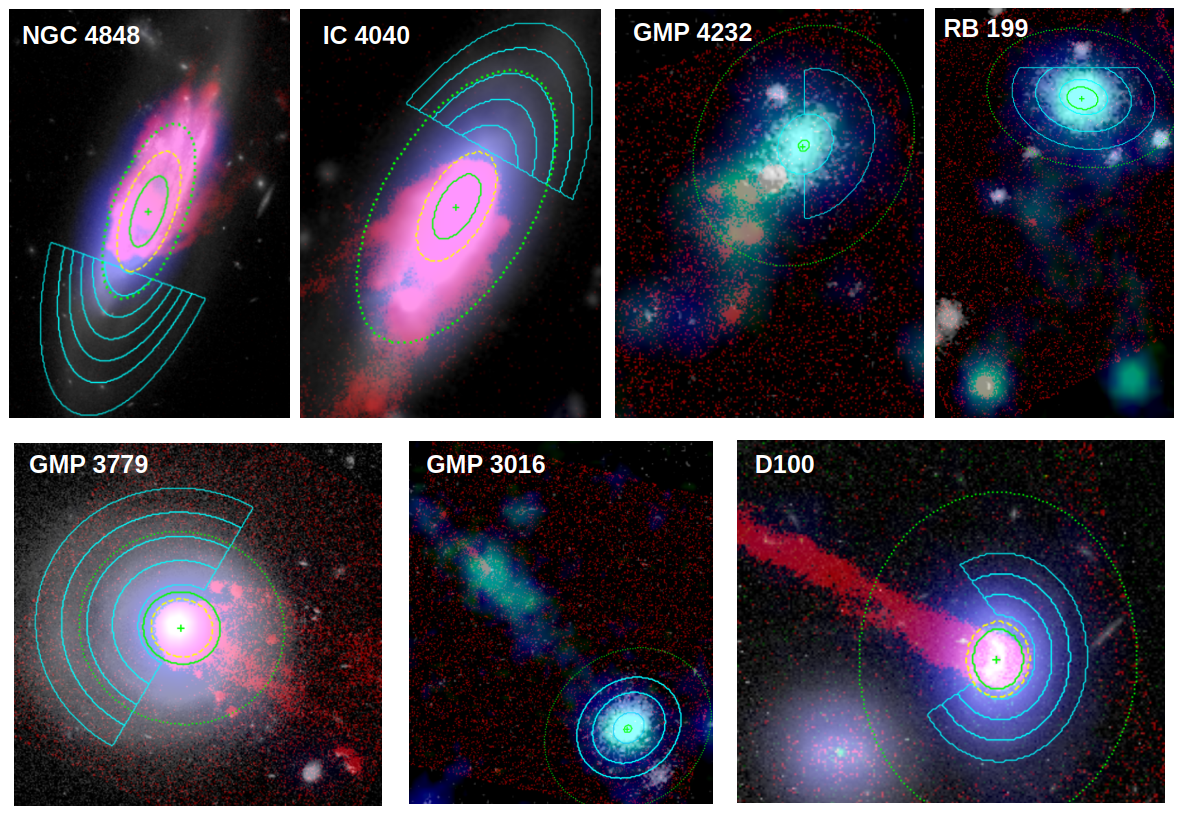}
    \includegraphics[width=1.\linewidth]{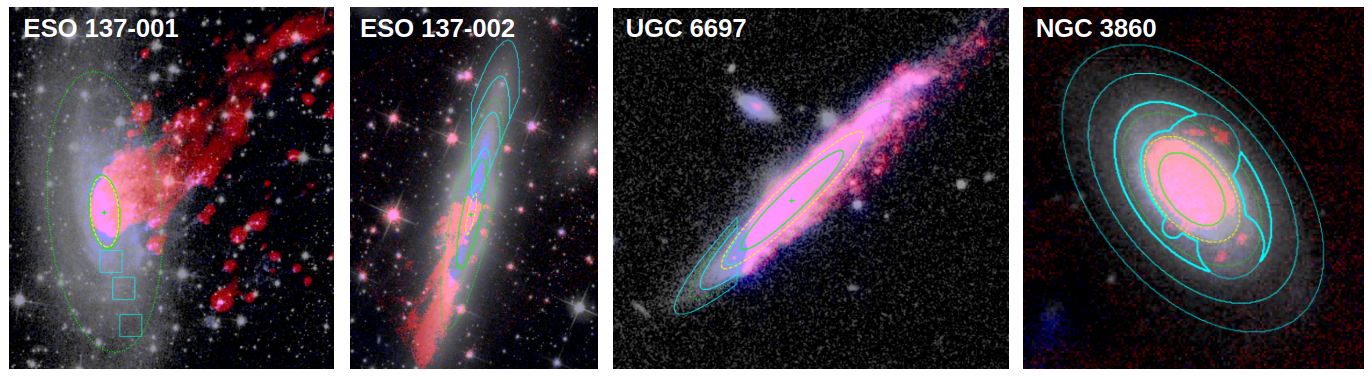}
    \caption{False color images of the galaxies of the sample (Table \ref{tab:galaxy_properties}) from Coma, Norma and A1367 clusters. The H$\alpha$ emission is shown in red, the GALEX NUV emission is shown in blue, while the bulk old stellar component (HST F814W if present, or SDSS $r-$band) is shown in greyscale. For RB 199, GMP 4232 and GMP 3016, the GALEX FUV emission is shown in green.
    The green solid and dotted lines represent the disc scale radius and a constant 8 kpc radius, respectively. The yellow dashed lines represent the truncation radius, as defined in Sect. \ref{sec:identification_regions_of_interes}. The regions we study in this work are shown in cyan. In the case of Norma, due to the proximity to the Galactic plane, the image is contaminated by foreground stars. To mitigate their contamination, for ESO 137-001, we used rectangular regions rather than elliptical ones.}
    \label{fig:Coma}
\end{figure*}

\subsection{Identification of the regions of interest}\label{sec:identification_regions_of_interes}

The regions used for the spectro-photometric analysis cover only quenched areas and split the area into approximately constant bins in galactocentric radius. Apertures are shaped as elliptical annuli centered on the galaxy nucleus, with their ellipticity and position angle matched to the optical $r$-band isophotes (or the HST F814W filter for Norma galaxies) at surface brightness levels of 24 mag arcsec$^{-2}$. To reach the outermost parts of the stellar disc, and thanks to the depth of our dataset, the outer radius of the outermost aperture is set to the 25 mag arcsec$^{-2}$ isophote.

Although higher spatial resolution could be achieved by further subdividing our apertures, sufficiently large apertures are required to
achieve signal-to-noise ratios greater than 10 in the UV and optical photometry and in each resolution element of the spectra. We also make sure that different annuli are spatially resolved at the resolution of the GALEX images ($\approx 5$ arcsec). The Herschel/PACS images are visually inspected, finding no detectable emission in any region. Therefore we treat the FIR points as upper limits. The 37 selected regions that constitute our final sample are shown in Fig. \ref{fig:Coma}.

Photometric measurements are derived from the multi-wavelength dataset by projecting each region over all the available bands. 
In most images, the dominant source of uncertainty is the sky background. To estimate both the background level and its uncertainty while accounting for spatially correlated noise over the extraction area, each extraction mask is randomly shifted within the field of view 1000 times, excluding positions overlapping the target galaxy or bright stars. The distribution of fluxes measured in these shifted apertures provides an empirical estimate of the background mean value and its associated uncertainty. This method naturally incorporates the effects of correlated noise induced by the PSF, detector characteristics, and resampling during data processing, which would otherwise invalidate assumptions of pixel independence. UV data (GALEX and UVIT) operate in a nearly photon-limited regime due to low count rates. Nonetheless we model and subtract the sky background from the apertures used to extract the photometry. Photometric measurements with a signal-to-noise ratio below 3 are treated as upper limits. In addition, the three FIR data points are set as upper limits in all regions, as described above. 

For each galaxy, we define the truncation radius as the semi-major axis of the elliptical isophote at which the \Ha{} surface brightness falls below $10^{-18}$~erg~s$^{-1}$~cm$^{-2}$~arcsec$^{-2}$. The ellipses have the same axis ratios and orientations adopted for the regions of interest and are shown in yellow in Fig.~\ref{fig:Coma}. Because ram pressure can produce asymmetric \Ha{} morphologies and extraplanar emission projected onto the disc, we identify the truncation radius along the major axis on the leading side of the RPS wind.

\section{Stellar population models}\label{Sec:Model}

To constrain the RPS episode through stellar population analysis, we adopt a parametric SFH that explicitly accounts for the suppression of star formation due to gas removal. Since our analysis aims at reconstructing the quenching histories in different galactocentric regions, the secular (unperturbed) SFH must include an explicit radial dependence, requiring a multi-zone formulation.

Following \citet{Boselli_2016_quenching_timescales} and \citet{Fossati_2018}, we adopt an unperturbed, radially resolved reference model plus a two-parameter quenching event. The SFH is defined as:
\begin{equation}
\label{eq:truncated_sfh}
\mathrm{SFR}(t)=
\begin{cases}
\mathrm{SFR}_{\mathrm{unpert.}}(t) & t < t_q \\
\mathrm{SFR}_{\mathrm{unpert.}}(t_q)\times
\exp\!\left[-\dfrac{t - t_q}{\tau_Q}\right] & t \geq t_q
\end{cases}
\end{equation}
where $t_q$ is the cosmological time at which the quenching event started, and $\tau_Q$ is the characteristic quenching timescale. We define the look-back time at which quenching begins as $Q_{\mathrm{AGE}} \equiv 13.5\ \text{Gyr}\ -t_q$.

The underlying unperturbed SFHs are taken from the multi-zone disc evolution model developed by \citet{Boissier_1999,Boissier_2000} and later updated by \citet{Buat_2008} and \citet{Munoz_Mateos_2011}. Initially calibrated on the Milky Way, this framework was generalized to other disc galaxies using the analytical scaling relations of \citet{MMW}. In this model, the SFH of a disc galaxy depends only on the galactocentric radius and is fully specified by two global parameters: the circular velocity of the disc $V_\text{rot}$ and the dimensionless spin parameter $\lambda$. The circular velocity and the disc scale radius $R_d$ are related to the stellar mass through
\begin{equation}\label{eq:V_scaling_MW}
\frac{V_\text{rot}}{220\,\mathrm{km\,s^{-1}}} =\left( \frac{M_*}{5 \times 10^{10}M_\odot} \right)^{1/3}
\end{equation}
\begin{equation}\label{eq:scale_radius_scaling_MW}
\frac{R_d}{2.6\,\mathrm{kpc}} = \frac{\lambda}{0.03} \cdot \frac{V_\text{rot}}{220\,\mathrm{km\,s^{-1}}}
\end{equation}
Given the spin parameter is nearly mass independent \citep{Maccio_2007} we fix it to $\lambda = 0.05$. This value is also consistent with those used in \cite{Boselli_2016_quenching_timescales} and \cite{Fossati_2018}.
The values of the circular velocity and of the scale radius obtained using these equations are reported in Table \ref{tab:galaxy_properties}.

The stripping event is described by two free parameters, $Q_{\mathrm{AGE}}$ and $\tau_Q$. This phenomenological parametrization is preferred over more physically detailed models in order to minimize the number of free parameters and limit degeneracies in the inferred quenching histories. One caveat of our choice is that we assume the quenching to be axisymmetric, thus neglecting the effects of disc rotation during the stripping and quenching event \citep{Sparre_2024}.  A more complex model including a burst of star formation induced by ram pressure and preceding the quenching phase is presented in Sect. \ref{sec:burst}.

From the unperturbed SFHs, we construct a grid of truncated models spanning $Q_{\mathrm{AGE}} = 10-3000$ Myr and $\tau_Q = 2-2000$ Myr with a logarithmically spaced bin size.
Given this SFH grid, spectral models are generated using the CIGALE code \citep{CIGALE}, assuming a fixed formation age of 13.5 Gyr and a \citet{chabrier2003galactic} initial mass function (IMF). We adopt the updated high-resolution Bruzual \& Charlot models (“cb19”; \citealt{Sanchez_2022}), with a fixed metallicity $Z=0.017$ (equal to the solar value in these models). 

Models are fitted to the spectral and photometric data using our new code \textit{Pangal}\footnote{https://github.com/robertoinfurna/Pangal}, which builds upon the \textsc{MC-SPF} \citep{Fossati_2018} code. The likelihood space is sampled using the \textit{dynesty} \citep{speagle2020dynesty} implementation of the nested sampling algorithm. The code takes as input the photometry and the extracted MUSE spectra for each region. We limit the spectrum to the range 4800-7000\AA; at longer wavelengths, the spectra are more significantly contaminated by OH line residuals and offer fewer constraints on the age of the stellar populations. 
The code interpolates a 4-dimensional grid ($V_\text{rot}$, $R$, $Q_{\mathrm{AGE}}$ and $\tau_Q$) of pre-computed models on the run and builds a composite spectrum from the stellar SED for each likelihood evaluation. We use flat priors spanning the full grid range on the two quenching parameters. Instead, we use narrower priors on $V_\text{rot}$ and $R$. For each galaxy, $V_\text{rot}$ is assigned a uniform prior spanning $\pm 10\,\mathrm{km\,s^{-1}}$ around the fiducial value reported in Table~\ref{tab:galaxy_properties}, while $R$ is allowed to vary between the inner and outer radii of each region.

Nebular emission is added to the stellar SEDs during the fitting procedure. We use the nebular emission template of \citet{Byler_2017} with solar metallicity. This template has two free parameters, the ionization parameter $\mathcal{U}$ and the age of the SSP ionizing the gas. The intensity of the emission lines is set by the rate of ionizing photons $Q_H$, which, in turn, is set by the instantaneous SFR and the assumed IMF. 
We assume a \cite{calzetti2000} attenuation law with a single free parameter $A_V$ for the old stellar component (Age $\geq 10$ Myr), while the young stellar component (Age $<10$ Myr) and the associated nebular emission are attenuated by an additional term $A_{V,\mathrm{extra}}$, nominally treated as a free parameter but with a narrow Gaussian prior centred on $1.17\ A_V$ and a standard deviation of 0.01.
Dust emission is modelled using the \citet{DH02} templates, which assume that infrared emission arises from dust exposed to a distribution of radiation field intensities $U$, ranging from 0.3 to $10^5$, where $U=1$ is the value for the radiation field in the solar neighbourhood \citep{Dale_2001}. The dust mass distribution follows $dM_d(U) \propto U^{-\alpha} dU$, where $\alpha$ controls the relative contribution of high- and low-intensity radiation fields. Low values of $\alpha$ indicate dust predominantly heated by star-forming regions, while high values correspond to dust heated mainly by the diffuse interstellar light. This model is adopted to limit the number of free parameters relative to more complex prescriptions. 
MIR and FIR data are essential to break the dust–age degeneracy, as constraining the dust emission enables a robust estimate of the UV–optical attenuation. The final synthetic spectrum is then convolved with the filter transmission curves to generate model photometry, which is compared with the observed data.

Spectral fitting requires additional processing steps. The model spectrum is convolved with a Gaussian line-of-sight velocity distribution (LOSVD) with observed frame velocity $v_\mathrm{sys}$ and dispersion: $\sigma^2 = \sigma_\text{obs}^2-\sigma_\text{model}^2+ \sigma_v^2$, which accounts for the finite resolution of the model SEDs, the MUSE instrumental spectral resolution, and intrinsic line broadening. The nebular emission lines are modelled at the same systemic redshift as the stellar component, but with a free parameter for their velocity dispersion, $\sigma_\text{gas}$. The velocity convolution is performed using the FFT-based method described in \citet{Cappellari_2017, Cappellari_2022}. Residual differences in continuum shape between model and observed spectra, primarily due to uncertainties in the MUSE flux calibration, are corrected by multiplying the model SED by a seventh-order polynomial during the fitting procedure.
The total likelihood is computed as the sum of photometric and spectroscopic likelihood terms. To account for potential underestimation of spectroscopic uncertainties, which may otherwise dominate the total likelihood, we introduce a scaling parameter for the spectroscopic errors, and we treat it within the fitting framework.

\begin{figure}
    \centering
    \includegraphics[width=0.99\linewidth]{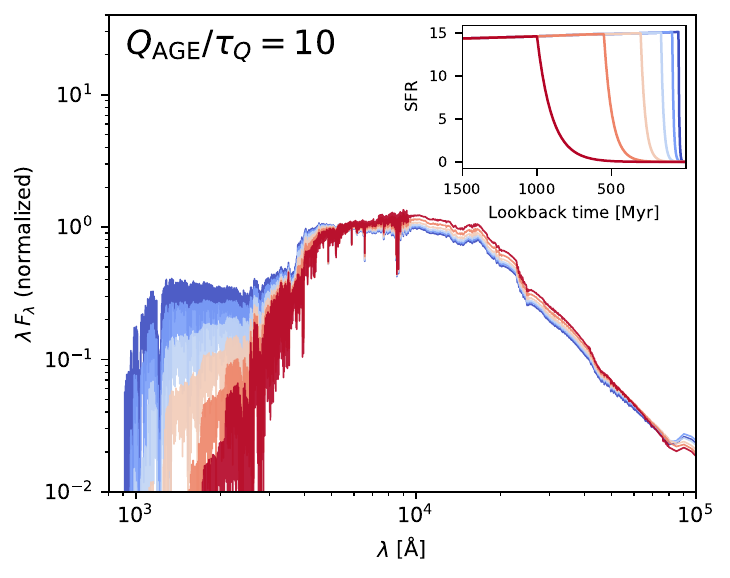}
    \includegraphics[width=0.99\linewidth]{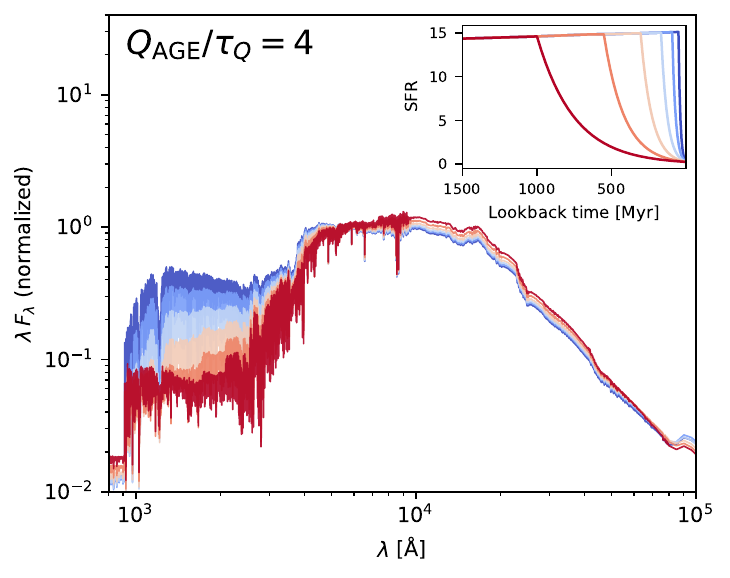}
    \caption{SED models generated with \textit{Pangal} as described in Section \ref{Sec:Model}. All models assume $V_\mathrm{rot}=160$ km s$^{-1}$, $R=8$ kpc, solar metallicity ($Z=0.017$), very low dust absorption ($A_V=0.01$), and fixed nebular gas parameters ($f_\text{esc}=0$, $\log \mathcal{U}=-3$, gas age $= 5$ Myr, $\sigma_\text{gas}=30$ km s$^{-1}$). All SEDs are normalised to the flux at $5500\AA$. The SFR curves in the insets are in units of $M_\odot \,\text{pc}^{-2}\, \text{Gyr}^{-1}$.} 
    \label{fig:models}
\end{figure}

\section{Mock tests of the fitting code}\label{Sec:Simulations}
In order to test our code and investigate the potential degeneracies arising from the large number of free parameters involved, we generated synthetic data from known quenching parameters, added realistic observational uncertainties, and then ran the fitting procedure to assess the extent to which we are able to recover the input model parameters. 

\begin{figure*}
    \centering
    \includegraphics[width=0.85\linewidth]{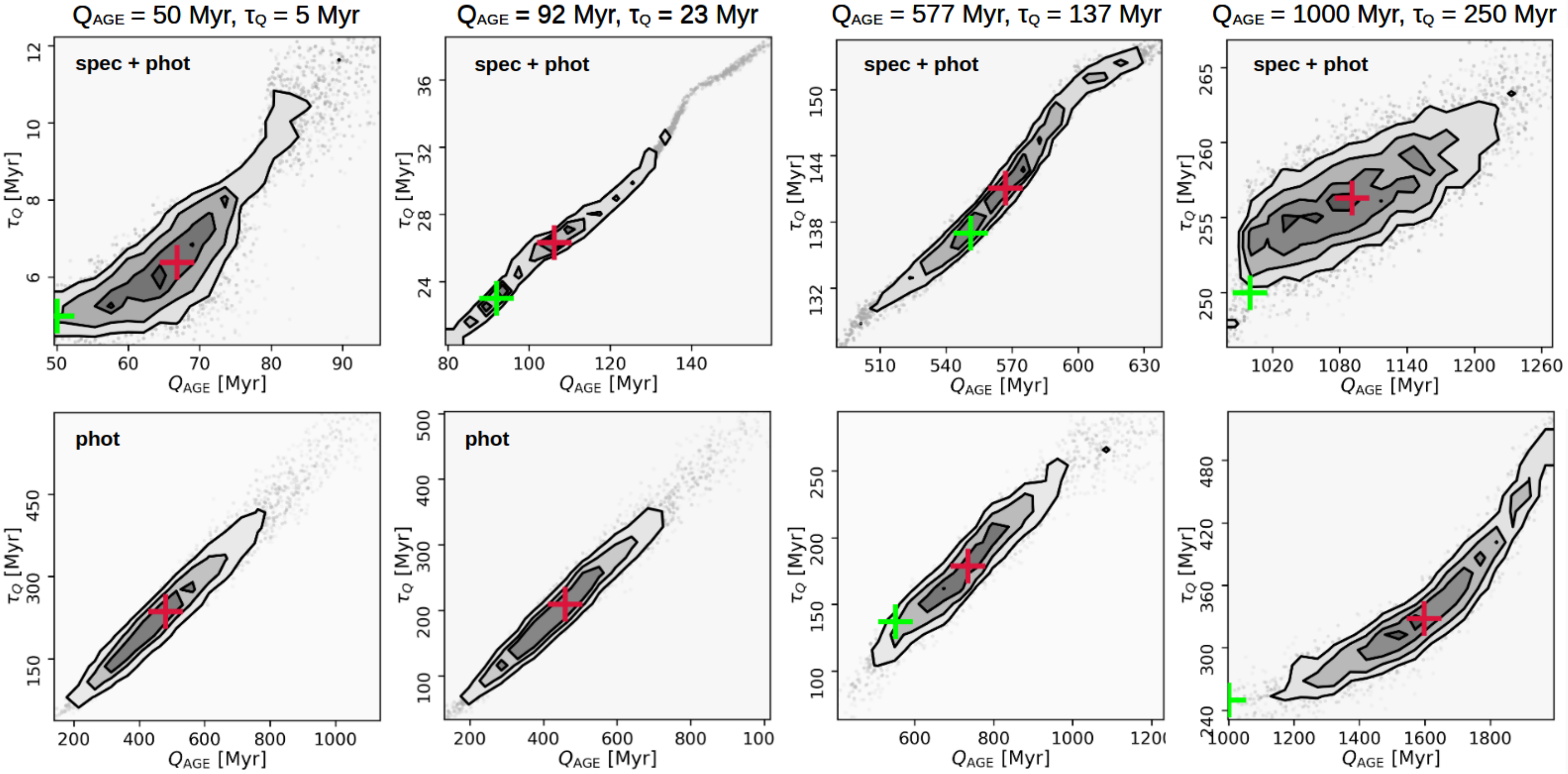}
    \caption{Posterior distributions in the $Q_{\rm AGE}$--$\tau_Q$ plane obtained from representative mock tests with different input quenching histories. The top row shows the results obtained by fitting the full mock dataset (photometry + spectroscopy), while the bottom row shows the corresponding fits using photometry alone. The input parameters are marked by the green crosses, and the posterior medians by the red crosses. The posterior distributions exhibit the degeneracy between $Q_{\rm AGE}$ and $\tau_Q$, with the data primarily constraining their ratio rather than the individual parameters. Including the spectroscopic information significantly reduces the allowed parameter space, whereas photometry alone leads to much broader posteriors and can result in large systematic offsets from the input values.}
    \label{fig:tests_cornerplots}
\end{figure*}

\begin{figure*}
    \centering
    \includegraphics[width=0.82\linewidth]{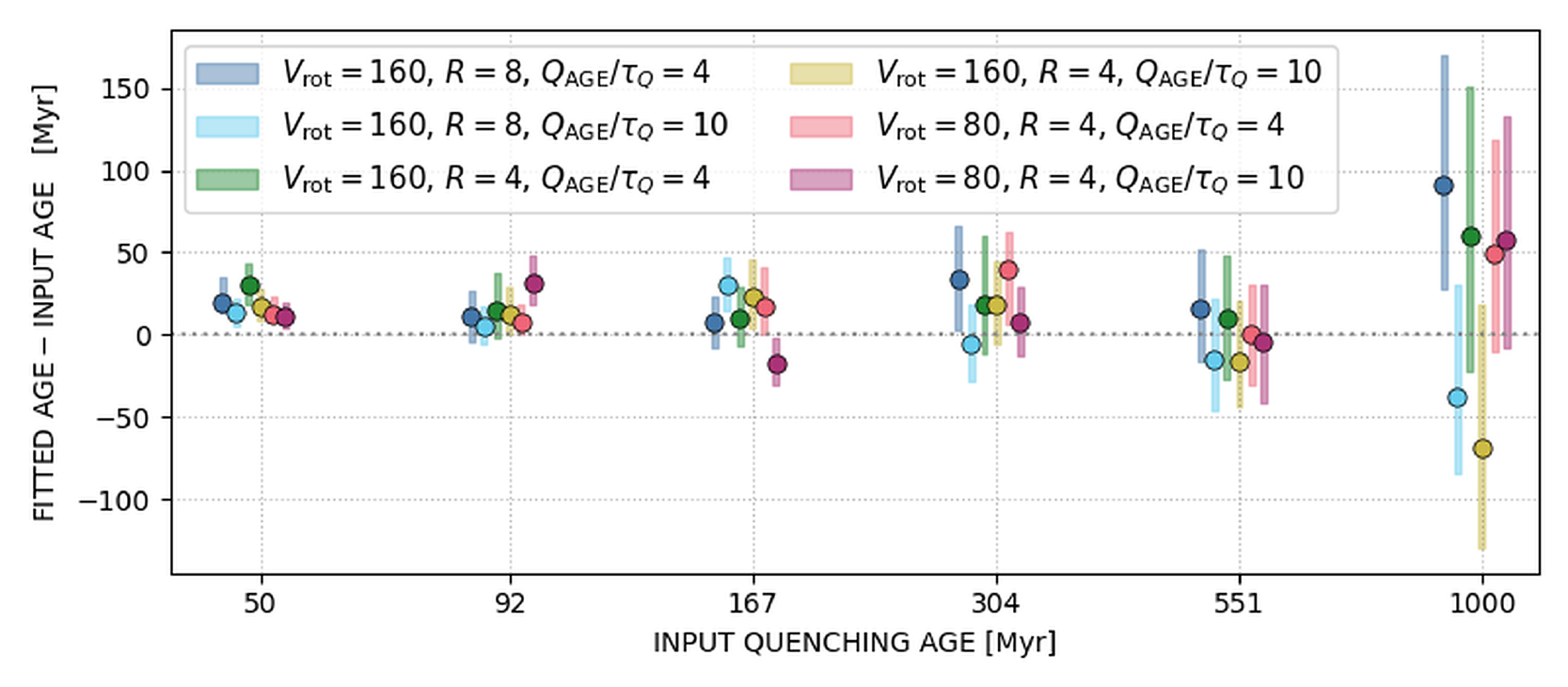}
    \caption{Difference between the fitted and input quenching age, $Q_{\rm AGE}$, as a function of the input quenching age for the mock tests. Input ages are logarithmically spaced in 6 bins from 50 to 1000 Myr. Different colours identify six different combinations of $V_{\rm rot}$, $R$, and $Q_{\rm AGE}/\tau_Q$. Symbols show the median of the posterior distributions, while the error bars correspond to the 16$^\mathrm{th}$ and 84$^\mathrm{th}$ percentiles. The recovered quenching ages show very good consistency with the input values. A small positive bias is present at the youngest ages, probably due to the similarity in the likelihood of very young models, given the SNR of our observational datasets.}
    \label{fig:tests}
\end{figure*}

Mock photometric and spectroscopic data were generated as follows: for the photometry, the model spectra were convolved with the transmission curves of the instruments described in Section~\ref{Sec:Data}. We then added both statistical and calibration uncertainties. The statistical uncertainties were assumed to be band-dependent, ranging from 3\% in the optical SDSS bands up to 15\% in the UV bands, while an additional 3\% calibration uncertainty was included for all bands. Random Poisson (statistical) and Gaussian (calibration) noise terms were applied to the photometric fluxes using these uncertainties. 
For the mock spectra, the models were first re-sampled onto the same wavelength grid and brought to the MUSE spectral resolution. Gaussian noise corresponding to a constant (1\%) fractional uncertainty, typical of MUSE spectra, was then added at each wavelength element. 

We run two sets of simulations, the first for a galaxy with $V_\text{rot} = 160$ km s$^{-1}$ and the second for a galaxy with $V_\text{rot} = 80$ km s$^{-1}$, in both cases with solar metallicity ($Z=0.017$), low dust extinction ($A_V=0.01$), fixed nebular gas parameters ($\log \mathcal{U}=-3$, gas age $= 5$ Myr, $\sigma_\text{gas}=30$ km s$^{-1}$).
In the first case, we explored two different galactocentric radii, 4 kpc and 8 kpc; in the second case, only 4 kpc. We generated mock spectra with input quenching ages, $Q_{\rm AGE}$, uniformly distributed in logarithmic space between 50 and 1000 Myr. For each model, the quenching timescale was set to either $\tau_Q = Q_{\rm AGE}/4$ or $\tau_Q = Q_{\rm AGE}/10$. The resulting spectral energy distributions (SEDs) for the $V_\mathrm{rot}=160$ km s$^{-1}$, $R=8$ kpc models are shown in Fig.~\ref{fig:models}. It is worth noting that models with short quenching timescales may still exhibit weak nebular emission lines. In particular, all models with $\tau_Q = Q_{\rm AGE}/4$ retain H$\alpha$ in emission, while H$\beta$ is dominated by absorption. Although the observed galaxies in our sample do not show detectable emission lines, it is nevertheless important to include nebular emission in the modeling, such that the models have to reproduce the emission line properties observed. 
This also illustrates why photometry alone is insufficient to constrain the quenching history. Broadband photometry cannot rule out models that exhibit significant nebular emission. For all the fits we set, as a prior, that the disc velocity can range $\pm 10$ km s$^{-1}$ around the nominal value, and $R\pm 1$~kpc around the nominal value. 

All the obtained posterior distributions reveal a strong degeneracy in the $Q_{\rm AGE}-\tau_Q$ plane (see Fig. \ref{fig:tests_cornerplots}), consistent with the results of \cite{Fossati_2018} on real data. The observations constrain a family of quenching histories, i.e., an older quenching event with a long decay timescale can reproduce spectra similar to those produced by a younger quenching event with a shorter decay timescale. Consequently, the posterior distribution is highly elongated showing that we primarily constrain the ratio of those parameters rather than their individual values. The remaining parameters are generally well constrained and show little evidence of significant degeneracies. The parameters $V_\mathrm{rot}$ and $R$ are in principle degenerate with the quenching parameters, as they determine the underlying unperturbed SFH. However, the narrow priors adopted limit the impact of this degeneracy.

\begin{figure*}[!t]
    \centering
    \includegraphics[width=0.9\linewidth]{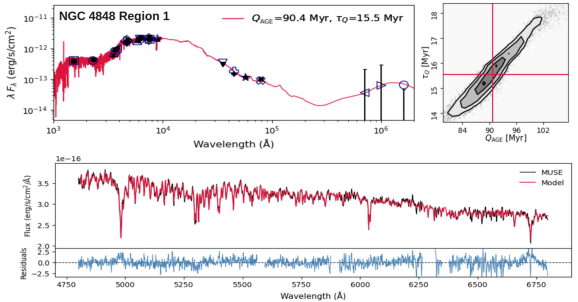}
    \caption{Results of the fitting procedure for Region 1 of NGC 4848. Top-left panel: the photometric data points (black symbols) are superimposed on the best fit spectral model. The synthesized photometric points are shown as open symbols outlined in blue. The FIR data are upper limits. Bottom panel: MUSE spectrum (black solid), and best fit model (red). The spectral residuals, normalised to the data uncertainty, are shown below the spectrum. Top-right panel: the posterior distributions of $Q_\mathrm{AGE}$ versus $\tau_Q$. The red cross shows the median values of the marginalized posterior distributions of each parameter.}
    \label{fig:fit_example_one_region}
\end{figure*}

\begin{figure*}
    \centering
    \includegraphics[width=1.\linewidth]{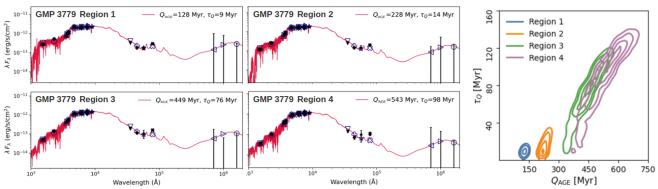}
    \caption{Four regions of GMP 3779. Only the photometric data are shown, and points and lines are color coded as in Figure \ref{fig:fit_example_one_region}. The progressive aging of the stellar populations from Region 1 (innermost) to Region 4 (outermost) is made clear by the progressive decrease in the UV flux. The corresponding posterior distributions in the $Q_\mathrm{AGE}$-$\tau_Q$ parameter space are shown in the rightmost panel.}
    \label{fig:gmp3779_allregions}
\end{figure*}

Results are presented in Fig.~\ref{fig:tests}. The markers indicate the median values of the posterior distributions, while the error bars span the 16th and 84th percentiles. We find that, at low input quenching ages, the fitted median values tend to be higher than the true input values, while this effect becomes less evident at larger ages. This is due to the fact that the model grid is limited at young ages and short quenching timescales. As a consequence, the sampler can explore regions of parameter space with larger ages and longer timescales, but not regions with younger ages and shorter timescales. The posterior median is therefore biased toward higher values. We also performed an additional test by fitting the mock data with a very high signal-to-noise ratio of 1000 in the FUV and NUV bands. In this case, the fitting procedure was able to recover the input values without significant deviations. This shows that the uncertainty in the UV photometric points is one of the main sources of bias in the recovered quenching parameters. Overall, the fitting code proves to be robust, since the typical offset of about 25 Myr is small compared to the typical timescales explored in this work, and we regard this value as the minimum time resolution recoverable with our method. 
The mock tests also demonstrate that this analysis cannot be performed reliably using photometry alone. In that case, the posterior distributions become much broader and systematic offsets of up to $\sim500$ Myr can occur. This happens because models with old quenching ages and long quenching timescales would have nebular emission. However, they cannot be ruled out using broadband photometry alone and become excluded only when the spectroscopic information is included in the fit.

\section{Results}\label{Sec:Results}

In order to accurately reconstruct  the radial quenching history for the 11 galaxies of the sample, we ﬁt the photometry and spectroscopy of all the regions, following the techniques described in Section~\ref{Sec:Model}. 
Figures~\ref{fig:fit_example_one_region}-\ref{fig:gmp3779_allregions}
show examples of the output of the \textit{Pangal} code applied to a single region of NGC 4848 and all regions of GMP 3779. In red, we show the median models drawn from the posterior distributions.

\begin{figure*}[!t]
    \centering
    \includegraphics[width=0.95\linewidth]{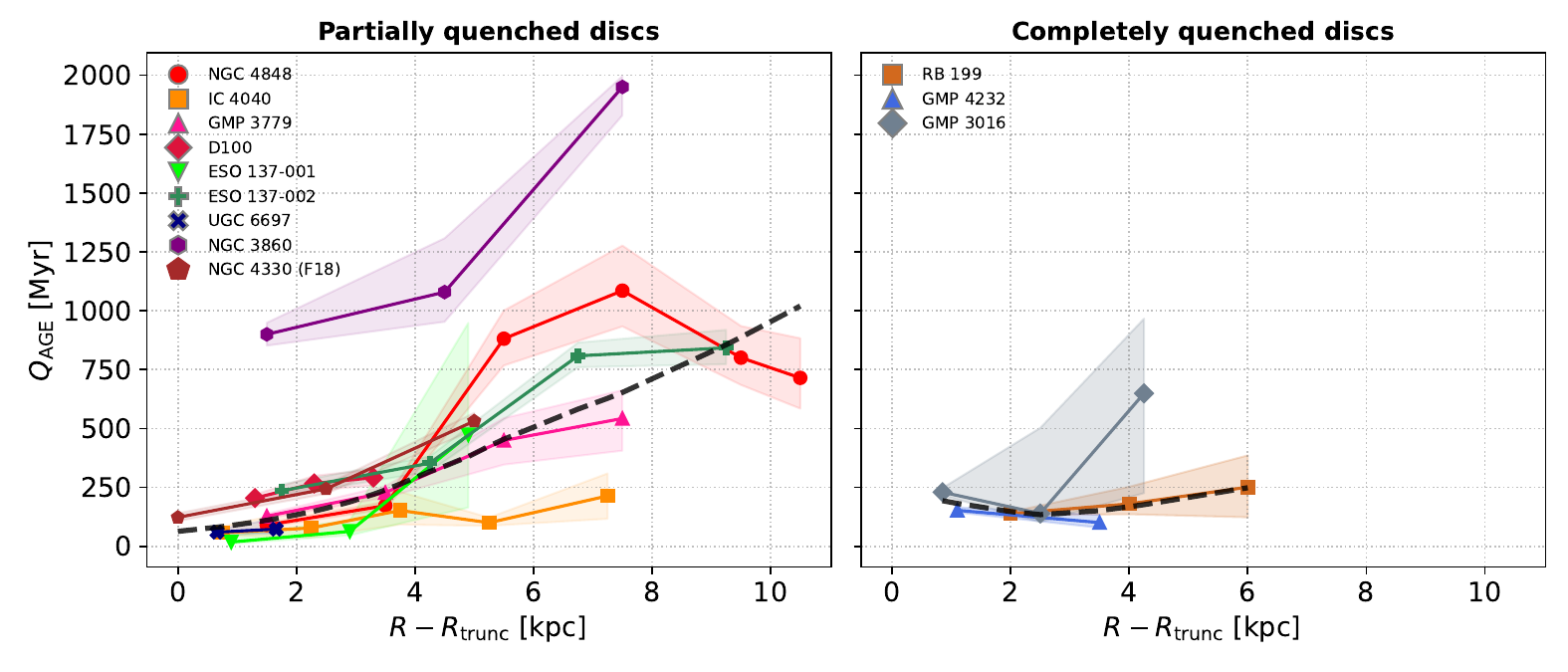}
    \caption{Radial trends in the quenching age, $Q_\mathrm{AGE}$, for the 11 galaxies in the sample. All galaxies are fitted assuming a fixed metallicity of $Z=0.017$. The points indicate the marginalized posterior medians, while the shaded regions span the interval between the 0.05 and 0.95 posterior quantiles. We also include literature values for NGC 4330 from \citet{Fossati_2018}, obtained using a similar analysis. The $x$-axis shows the distance of each region from the truncation radius of its host galaxy. The left panels show galaxies with a star-forming nuclear region and a stripped gas tail connected to the disc. Their truncation radii are defined in Section~\ref{sec:identification_regions_of_interes} and shown in yellow in Fig.~\ref{fig:Coma}. The right panels show galaxies without a star-forming nucleus, which appear to be fully quenched. For these galaxies, we adopt $R_\mathrm{trunc}=0$. They are also the three least massive galaxies in the sample. The black dashed lines represent the average of the data points and highlight the underlying trends.}
    \label{fig:results_final}
\end{figure*}

In all cases, FIR points are treated as upper limits; therefore, dust is not fully constrained, but we can rule out dust absorption higher than $A_V\approx0.2$. An upturn of the IRAC 8 $\mu$m point is a good indicator of the presence of dust, but most regions are consistent with no dust. The inclusion of nebular lines in the model is driven by the choice of having flexible models that can account for a small amount of residual star formation, especially in the regions close to star-forming regions in discs where the quenching is not complete over the disc extent. For the same reasons, our SFH model allows low values of $Q_\text{AGE}/\tau_Q$, where current SFR could still be important.

Fig.~\ref{fig:results_final} shows the quenching ages $Q_\mathrm{AGE}$ inferred from the retrievals at their distance from the truncation radii of each galaxy, as defined in Section~\ref{sec:identification_regions_of_interes} and shown in yellow in Fig.~\ref{fig:Coma}. The galactocentric radius of the finite-size regions, $R$, is defined as the midpoint of the inner and the outer radii.
We divided the galaxies in two subsamples: the ones with a star forming central region, which are still undergoing the stripping process; and the ones with no \Ha{} emission ($\lesssim10^{-18} \mathrm{erg}\,\mathrm{s}^{-1}\,\mathrm{cm}^{-2}\,\mathrm{arcsec}^{-2}$) which are fully quenched and therefore have $R_\text{trunc}=0$. Interestingly the galaxies of this second class, RB 199, GMP 4232 and GMP 3016, are the three least massive galaxies of the sample. Therefore we will refer the first class as intermediate and high-mass galaxies and the second class as low-mass galaxies (GMP 4232 and GMP 3016 in particular are classified as dwarf galaxies).

We observe a clear trend in the high- and intermediate-mass galaxies: the quenching age increases monotonically with galactocentric distance. There are only two exceptions, namely the two outermost regions of NGC 4848. However, these regions have lower S/N than the inner regions, making the inferred quenching ages less significant. 
The black dashed line in the figure describes the overall trend. Most of the galaxies cluster around this value, with the exception of NGC 3860, which has much longer quenching ages than any other galaxy. Indeed, this galaxy lacks an \Ha\ tail, indicating that the amount of mass loss due to stripping is currently low. Its older quenching times are likely related to its selection as the only sample galaxy lacking gas in the tail. We conclude that the galaxy is currently in a post-peak stripping phase, in which the current stripping rate is low, and the gas truncation radius is not changing much over time.

The trends are less clear in the low-mass galaxies, particularly for the two dwarf galaxies GMP 4232 and GMP 3016. These two galaxies are less extended ($\sim 15-16$ arcsec in diameter) and the S/N ratio of both integrated spectrum and photometry is lower compared to the high-mass galaxies, making the analysis more difficult. In RB 199 and GMP 4232 we find short quenching times and nearly flat profiles, also given the larger uncertainties of our analysis for these faint objects. This is highlighted by the case of GMP 3016, which in the outermost region presents a quenching age much longer than the inner regions, but with very high uncertainty.

The results for the full sample as a function of the projected galactocentric radius $R$ of each region are presented in Fig.~\ref{fig:results_all_galaxies}. We report the quenching parameters $Q_\text{AGE}$ (lookback time), $\tau_Q$ (e-folding timescale) and $Q_\text{AGE}/\tau_Q$ (number of $e$-foldings of the star formation suppression). Neither $\tau_Q$ nor $Q_{\mathrm{AGE}}/\tau_Q$ show a significant radial gradient. Most galaxies have $\tau_Q \lesssim 100$~Myr, although the uncertainties are often substantial. Similarly, the radial profiles of $Q_{\mathrm{AGE}}/\tau_Q$ are generally flat, with values for most galaxies clustering around $\sim 5$.

\subsection{Effects of the inclusion of a burst in the SFHs}\label{sec:burst}

Observational campaigns revealed that ram-pressure can induce a transient enhancement of star formation. Galaxies undergoing strong RPS often display asymmetric \Ha{} morphologies, including one-sided, ring-like, or crescent-shaped distributions tracing the interaction between the interstellar and intracluster media \citep{Gavazzi_2001,Koopman_2004, GASP_2017}. Stellar population studies have found very strong Balmer lines in the recently stripped discs of Virgo galaxies, consistent with starbursts near the time of quenching \citep{Crowl_2006,Kenney_2014}.  In several systems, enhanced star formation is observed preferentially along the leading edge of the stellar disc, where the ram pressure is expected to be strongest, while the downstream regions are progressively depleted of gas, \citep[e.g. IC 3476 in Virgo][]{Boselli_2021}.  Numerical simulations have shown that this enhancement is primarily driven by the compression of the interstellar medium and by ram-pressure-induced gas inflows towards the central regions, which temporarily increase the molecular gas content and the dense gas fraction available for star formation \citep{Kapferer_2009, Bekki_2014, Steinhauser_2016, Zhu_2024}. 

\begin{figure*}[!t]
    \centering
    \includegraphics[width=0.94\linewidth]{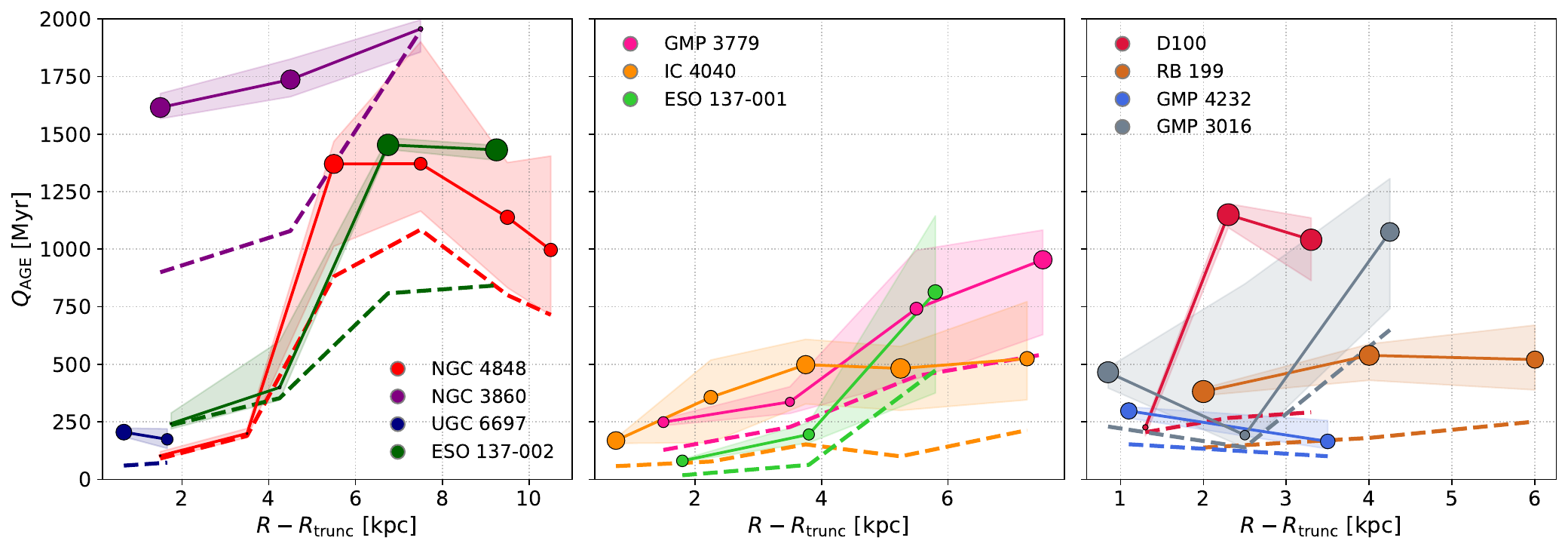}
    \caption{Radial profiles of the quenching age, $Q_{\rm AGE}$, obtained with (solid lines and filled circles) and without (dashed lines) the burst model. The three panels group galaxies in bins of decreasing stellar mass, from high (left) to low (right). Shaded regions indicate the 5th--95th percentile confidence intervals of the burst-model posterior distributions. The size of each marker is proportional to the inferred burst mass fraction ranging from 19\% (larger points) to 1\% (smaller points).}
    \label{fig:results_burst_all}
\end{figure*}

Motivated by these findings, we introduced a burst component into our quenched SFH model. The burst is modelled as an episode of enhanced star formation lasting 10 Myr immediately preceding the onset of the exponential quenching phase. While the choice of the burst duration is arbitrary, a 10 Myr duration ensures reaching a steady state for the ionizing spectrum during the burst phase and before the onset of quenching and is similar to the burst parametrization used in \cite{Cramer_2019}. For each region, its strength is parameterized by $f_\text{burst}$, defined as the ratio between the stellar mass formed during the burst and the total stellar mass formed over the entire star formation history, excluding the burst. 


We repeated the Bayesian parameter retrieval performed for the models without a burst, introducing $f_\text{burst}$ as an additional free parameter. The burst fraction was allowed to vary between 0 and 0.2. The resulting constraints on the quenching ages are presented in Fig.~\ref{fig:results_burst_all}. 
Incorporating a starburst prior to truncation systematically increases the inferred quenching ages across all mass regimes. On average, the inclusion of a burst increases $Q_{\mathrm{AGE}}$ by $\sim 25\%$ compared to the baseline no-burst scenario. This systematic offset occurs because the elevated UV flux generated during the burst rejuvenates the stellar population, requiring a longer subsequent quenching phase to reproduce the SEDs and absorption features observed today. 

We compared the Bayesian evidences of the baseline and burst models to assess the significance of the star formation burst. The burst model is strongly favored (difference in log evidence $\Delta\ln Z>5$) in Regions 1 and 2 of NGC 3860, UGC 6697, ESO 137-001, and RB 199; Regions 3 and 4 of ESO 137-002; Region 1 of GMP 3779, GMP 4232, and GMP 3016; and all regions of IC 4040. The burst is not required, in a Bayesian sense, in any of the regions of NGC 4848 and D100. It appears that the burst, when required, is often located in the innermost regions of these galaxies, close to the truncation radius, although we cannot completely rule out that the burst is more difficult to identify in the outer and older regions of the target galaxies. 

The relative radial gradients in $Q_{\mathrm{AGE}}$ remain unchanged with and without the inclusion of the burst. Furthermore, we do not find any significant correlation among the inferred burst strength ($f_{\mathrm{burst}}$), the relative increase in quenching time, and the host galaxy stellar mass. The fitted burst fractions are relatively small ($f_{\mathrm{burst}}$ is on average 0.09 reaching up to 0.19), indicating that while RPS-induced ISM compression can trigger localized starbursts, it does not dominate the overall stellar mass assembly. Consequently, our primary conclusions regarding mass-dependent quenching timescales remain robust against the choice of the SFH parametrization.

\section{Discussion}\label{Sec:Discussion}

We have built a sample of galaxies undergoing RPS in well-known local massive clusters, characterized by high-quality data. We now discuss the implications of the results drawn from our full sample, while Appendix \ref{sec:discussion_individual} is dedicated to a more in-depth analysis of the individual objects.

The classic condition for efficient stripping proposed by \citet{GunnGott1972} yields an expression for the stripping radius. 
Empirically, however, only the projected distance from the cluster's center and the line-of-sight velocity relative to the cluster can be constrained observationally. The infall velocity $V$, and the three-dimensional position within the cluster - and consequently the intracluster medium density $\rho_\text{ICM}$ - are difficult to estimate. The stellar mass and \HI{} gas, both observable quantities, are among the critical regulators of RPS. For a given $V$ and $\rho_\text{ICM}$, more massive galaxies have deeper potential wells, which should allow them to retain their gas more effectively. 

Our results are broadly consistent with this scenario: intermediate to high-mass galaxies that are partially quenched (see Fig.~\ref{fig:results_final}, left panel) demonstrate that RPS can lead to a gradual, outside-in quenching of star formation. As a galaxy accelerates towards the cluster, the combination of modest infall velocity and low ambient ICM density can efficiently strip only the outer edges of the HI discs, while the innermost regions of the galaxies retain their gas. Only after several hundred Myr do the galaxies approach denser cluster regions, and a more intense ram pressure strips the gas within the scale radius.
Given the typical cluster velocity dispersions of $\sim1000$ km s$^{-1}$, a galaxy can travel approximately 1~Mpc ($\sim 2/3$ of the virial radius) in 1~Gyr, a quantity similar to the time offset between the inner and outer stellar population ages in these massive galaxies. 
Conversely, these trends are less pronounced, and nearly disappear in lower-mass galaxies that are completely quenched  (see Fig.~\ref{fig:results_final}, right panel), suggesting that their shallower potential wells make the gas more subject to stripping, leading to a rapid and complete quenching across the disc, as also reported by \cite{Boselli_2008_dwarf_ellipticals}.

\begin{figure}
    \centering
    \includegraphics[width=1\linewidth]{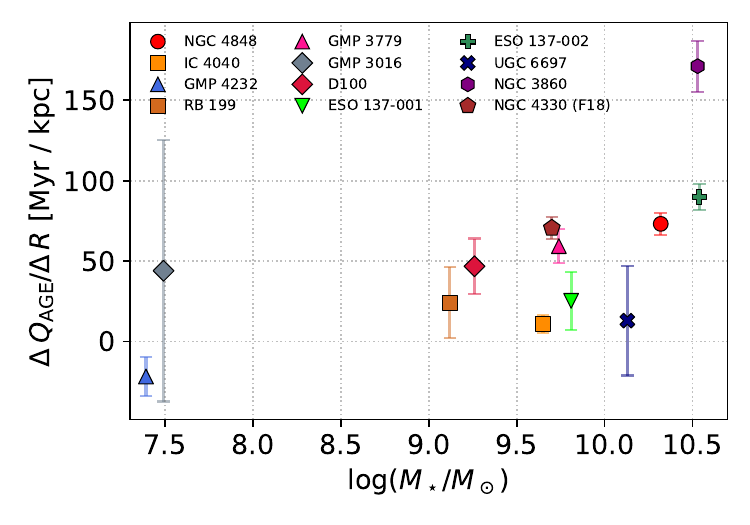}
    \caption{The quenching propagation rate, as parametrized by the gradient $\Delta Q_\mathrm{AGE}/\Delta R$, as a function of galaxy stellar mass. The gradients are derived by fitting a linear model to the radial $Q_\mathrm{AGE}$ profiles, and their uncertainties correspond to those associated with the fitted slopes.}
    \label{fig:stellar_mass_vs_qage}
\end{figure}

We summarise these mass-dependent quenching age trends in Fig.~\ref{fig:stellar_mass_vs_qage}, which shows a moderately strong trend with the stellar mass, despite a significant scatter between different objects at fixed stellar mass. To quantify the existence of a trend we use the Spearman rank correlation coefficient, obtaining a 1.5\% probability of the null hypothesis (no correlation). The scatter of individual data points is unavoidable, given that the RPS-driven quenching is a function not only of the satellite galaxy stellar mass but also of other variables, including the orbit of the galaxy within the cluster, the morphology of the hot ICM in individual clusters, the angle between the galaxy and the RPS wind, and the gas distribution in the galaxy, all of which change with time. Our sample is too small to reliably study the effects of additional variables. Larger and homogeneous samples are therefore required to quantify if and how this scatter is driven by those additional variables and to further boost the statistics of our results in the dwarf regime. 

Our results are in good agreement with previous MUSE studies of cluster PSB and quenched galaxies, notably those from the GASP survey \citep{Vulcani_2020, Werle_2022}. Using non-parametric spectral modeling with \textsc{Sinopsis} \citep{Fritz_2017},  \citet{Vulcani_2020} found that star formation suppression in local cluster galaxies typically proceeds outside-in on short timescales (from a few tens to several hundred Myr), with several objects showing evidence of central star formation bursts prior to complete shutdown. Similarly, \citet{Werle_2022} demonstrated that abrupt star formation truncation driven by environmental stripping operates efficiently in cluster environments out to intermediate redshifts ($z \sim 0.3$--$0.4$). The short quenching e-folding timescales inferred for most of our regions, 
$\tau_Q \lesssim 100$~Myr, are also consistent with the radio excesses observed in rapidly quenching cluster galaxies \citep{Ignesti_2022,Edler_2024}. A decline in star formation on timescales comparable to the lifetime of radio-emitting cosmic-ray electrons causes the radio continuum to lag behind the current SFR, producing enhanced radio luminosities relative to standard radio--SFR relations.

Our parametric spectrophotometric analysis reinforces this paradigm while offering an extended and complementary perspective: having access to a large range in stellar mass ($\log M_*/\mathrm{M}_\odot \sim 7.5$--$10.5$), we find that the radial quenching gradient is, to first order, regulated by the depth of the gravitational potential well. The positive correlation of the quenching age with mass is in apparent tension with a recent work by \cite{Marasco2026}. These authors, using a simplified cluster evolutionary model, found an anti-correlation of the star formation quenching timescale with the stellar-to-cluster mass ratio. The samples and analysis methodologies are, however, so different that we cannot directly compare their results to ours. 
Indeed, a feature of our sample is that it focuses on galaxies undergoing stripping and only the lower mass galaxies are fully quenched, thus allowing us to observe a change in the radial trends as a function of mass and stripping stage. 

\section{Conclusions}

We have investigated the quenching histories of 11 ram-pressure stripped galaxies in the Coma, Norma, and A1367 clusters using spatially resolved MUSE spectroscopy combined with multi-wavelength photometry. By fitting parametric SFHs within a Bayesian framework, we derived the quenching history in 37 regions distributed across the galaxy discs. Our results can be summarised as follows:
\begin{itemize}
    \item The majority of galaxies (8/11) exhibit quenched outer stellar discs devoid of ongoing star formation, while stripping is likely still affecting the active inner gaseous discs. Only GMP 3016, GMP 4232, and RB 199 in Coma appear entirely quenched.
    \item In most intermediate- and high-mass galaxies, we find a clear outside-in progression of quenching, with star formation suppressed only $\sim50$--250 Myr ago in the inner regions currently undergoing stripping, whereas the outer discs were already quenched $\gtrsim500$ Myr ago. 
    NGC 3860, the only sample galaxy without a prominent gas tail, is an exception to these trends, being in a late stripping phase thus having older stellar ages ($\approx 700$ Myr older than the average of the other galaxies in the sample). This is consistent with the expected connection between the disc quenching time and recent gas stripping rate.
    \item Lower-mass galaxies show much weaker radial gradients, suggesting that ram-pressure stripping removes the gas reservoir on significantly shorter timescales and soon after the first onset of RPS.
\end{itemize}
These results provide direct observational evidence that the efficiency and spatial progression of ram-pressure stripping are strongly regulated by galaxy mass, with massive galaxies undergoing a prolonged outside-in quenching process, while dwarf galaxies can be stripped almost entirely in a few hundred Myr. 

\begin{acknowledgements}
MS, PaS, KH, NP and JT acknowledge support from the NSF grant 2407821. PJ and AI acknowledges support from the institutional project RVO:67985815 and the project 25-19512L of the Czech Science Foundation.
This work is based on data obtained as part of the Canada-France Imaging Survey, a CFHT large program of the National Research Council of Canada and the French Centre National de la Recherche Scientifique and part of the UNIONS Survey. Based on observations made with the NASA/ESA Hubble Space Telescope and the Galaxy Evolution Explorer mission, obtained from the MAST data archive at the Space Telescope Science Institute, which is operated by the Association of Universities for Research in Astronomy, Inc., under NASA contract NAS5–26555. This work is based in part on archival data obtained with the Spitzer Space Telescope, which was operated by the Jet Propulsion Laboratory, California Institute of Technology under a contract with NASA. Based in part on observations made with the Herschel telescope, an ESA space observatory with science instruments provided by European-led Principal Investigator consortia and with important participation from NASA. Based on observations made with the ESO/VLT MUSE instrument and with the Joint ALMA observatory. ALMA is a partnership of ESO (representing its member states), NSF (USA) and NINS (Japan), together with NRC (Canada), NSTC and ASIAA (Taiwan), and KASI (Republic of Korea), in cooperation with the Republic of Chile. This work makes use of Sloan Digital Sky Survey data. Funding for SDSS has been provided by the Alfred P. Sloan Foundation, the Heising-Simons Foundation, the National Science Foundation, and the Participating Institutions. This work uses data from the AstroSat mission of the Indian Space Research Organisation (ISRO), archived at the Indian Space Science Data Centre (ISSDC). Data products generated in this work can be obtained, upon reasonable request, by contacting the corresponding author. 

\end{acknowledgements}

\bibliography{bibliography}

\begin{appendix}




    

\section{Quenching times as a function of projected galactocentric radius}

\label{app:radial}

\begin{figure*}[t]
  \centering
\includegraphics[width=1.0\linewidth]{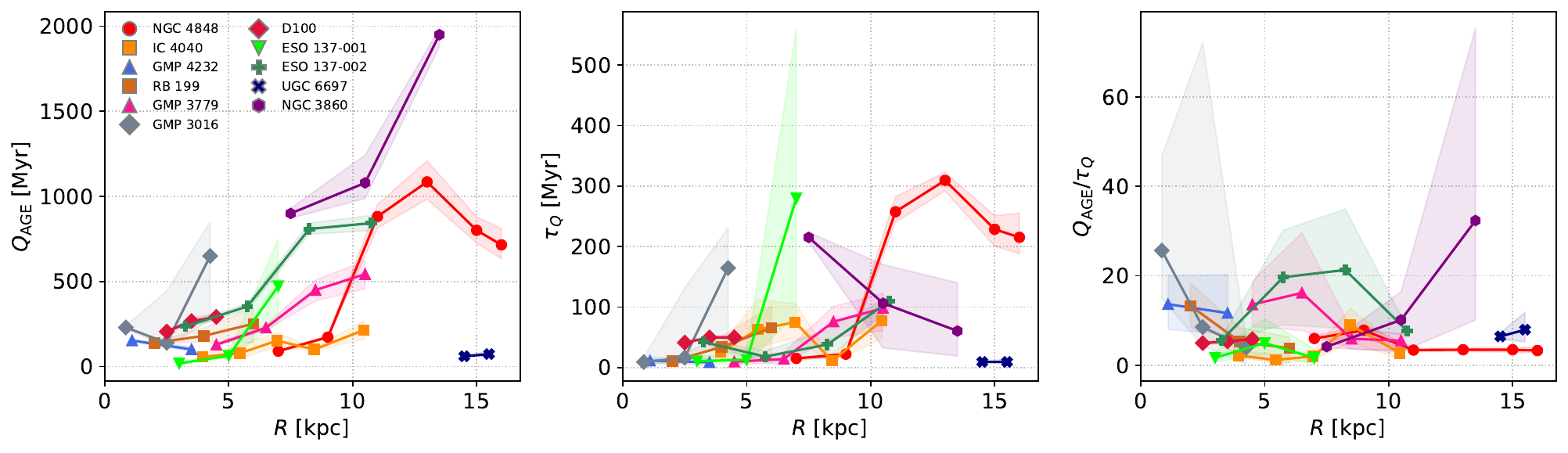}  
\caption{Best-fit quenching parameters for the 37 regions of the sample as a function of the projected galactocentric radius $R$: the
  quenching age $Q_{\rm AGE}$ (left), the e-folding timescale $\tau_Q$ (middle), and the number of e-foldings of the star formation suppression, $Q_{\rm AGE}/\tau_{Q}$ (right). All galaxies are fitted using the no-burst models with fixed metallicity $Z = 0.017$. The points correspond to the marginalized median values, while the shaded regions span the interval between the 0.05 and 0.95 posterior quantiles. No average trend line is drawn here, unlike in Fig.~\ref{fig:results_final}.}
  \label{fig:results_all_galaxies}
\end{figure*}

In Fig.~\ref{fig:results_all_galaxies} we show the quenching parameters inferred for all 37
regions of the sample as a function of the nominal projected galactocentric
radius $R$, defined as the midpoint between the inner and outer semi-major axes
of each aperture. This is similar to Fig.~\ref{fig:results_final}, where the same
posteriors are plotted against the distance from the truncation radius of each
galaxy. The left panel shows that the monotonic increase of $Q_{\rm AGE}$ with
radius is preserved in absolute radial units, but the scatter between different
systems at fixed $R$ is substantially larger, reaching $\sim 1$~Gyr at
$R \simeq 10$~kpc. This is expected, since $R$ mixes galaxies of very different
sizes and stripping stages: the dwarfs (GMP~4232, GMP~3016) only sample
$R \lesssim 4$~kpc, NGC~3860 reaches $Q_{\rm AGE} \simeq 2000$~Myr at
$R \simeq 13.5$~kpc, whereas the edge-on, recently stripped disc of UGC~6697 is
still quenching at $Q_{\rm AGE} \lesssim 100$~Myr out to $R \simeq 16$~kpc.
Normalising by $R_{\rm trunc}$, which encodes the instantaneous stripping front,
therefore removes most of this scatter and is a more physically meaningful
quantity for our investigation. The middle panel shows the e-folding timescale, $\tau_Q$. Most regions have $\tau_Q \lesssim 100$~Myr, with a mild trend with $R$, although this is probably driven by the correlation of this value with $Q_{\mathrm{AGE}}$. Instances where this is not the case are characterized by large uncertainties. The only significant exception is NGC~4848, whose outer regions have larger $\tau_Q$ values, reaching $\sim 300$~Myr. In this case also $Q_{\mathrm{AGE}}$ is high, giving a ratio of these two values consistent with other galaxies in the sample.  The right panel shows the number of e-foldings of the star formation
suppression, $Q_{\rm AGE}/\tau_{Q}$. Most regions have
$Q_{\rm AGE}/\tau_{Q} \simeq 2-10$ with no significant radial trend, confirming that star formation is essentially fully suppressed in all the quenched
apertures, and that the abruptness of the truncation is set by the local gas removal rather than by radius.

\section{The quenching history of individual galaxies}
\label{sec:discussion_individual}

We briefly discuss here how the quenching histories we derived in this work are related to the properties of the individual galaxies analysed and presented in Fig.~\ref{fig:Coma}.

\paragraph{Galaxies in Coma}

\begin{itemize}

    \item NGC 4848 is located at the NW periphery of the X-ray-emitting region of the Coma Cluster, at a projected distance of $\sim0.75$ Mpc ($r/r_{200}=0.49$) from the cluster center, and with a line-of-sight velocity similar to the cluster mean ($\Delta v \sim 275$ km s$^{-1}$). 
    A 65 kpc-long H$\alpha$ tail trails the galaxy and points in the direction opposite to the cluster core. This, together with kinematic considerations, strongly suggests a first-infall scenario \citep{Fossati_2012}. Our analysis suggests a gradual outside-in quenching scenario, in which the outermost disc regions ($R \gtrsim 11$ kpc) were depleted of gas approximately $750-1000$ Myr ago, while regions close to the truncation radius quenched more recently, around $\approx200$ Myr ago. The downturn in the quenching age at the largest radii could be physical but, given its modest statistical significance, we cannot rule out that it is partially caused by our assumptions (azimuthal symmetry, fixed metallicity). In the star-forming part of the disc, ram-pressure stripping might be triggering a starburst, which could explain the bright, blue, and clumpy ring of H\,{\sc ii} regions at $\sim 4$ kpc radius. 

    \item GMP 3779 is an intermediate-mass galaxy, south-west of the Coma cluster center at $r/r_{200}=0.35$. This galaxy shows an ionized gas tail pointing S-W. The downstream region is completely quenched and shows one of the best cases of outside-in RPS quenching with a nearly linear quenching profile with a gradient $\Delta Q_\mathrm{AGE}/\Delta R \approx 60$ Myr/kpc.
    
    \item IC 4040 (GMP 2559) is located north-east of the Coma cluster center at $r/r_{200}=0.16$. A bright \Ha{} filament, attached to the disc and pointing S-E extends $\sim 50$ kpc from the nucleus \citep{Yagi_2010}. The galaxy's nucleus is surrounded by bright, active star-forming regions that abruptly cut off on the northwestern side. The layout of the ionized gas points to a large-scale bow shock and a massive gas flow streaming southeast from the nucleus \citep{Yoshida_2012}, leaving the northwestern galactic disc entirely quenched. The quenching age profile is remarkably flat, and the outer disc of the galaxy ($R=2-7$ kpc) has been rapidly quenched over the last 250 Myr.
      
    \item D100 (Mrk 60, GMP 2910, CGCG 160-243) is an intermediate mass galaxy located south-east of the Coma cluster center at  $r/r_{200}=0.10$. 
    D100 also hosts a remarkably long and narrow \Ha\ tail \citep{Yagi_2010}, indicative of an advanced ram-pressure stripping stage in which most of the outer ISM has already been removed. The narrowness of the tail suggests that the stripping radius remained relatively stable at $\sim 1$ kpc over much of the stripping history \citep{Jachym_2017}. The star formation history of the disc was previously investigated by \citet{Cramer_2019} using HST/WFC3 colors. Their models assume a constant star formation rate over 12 Gyr followed by a sudden quenching. In addition, they considered a truncation-plus-burst scenario in which $2\%$ of the stellar mass formed at the onset of quenching, corresponding to a temporary increase in SFR by a factor of $\sim25$ over 10 Myr. They found that the colors are degenerate between simple truncation and truncation-plus-burst models, with burst models requiring older quenching ages systematically. Their analysis revealed a clear outside-in quenching pattern. 
    At $R\sim0.8$ kpc, their inferred quenching time is $\sim150\, (180)$ Myr for the $2 (5)\%$ burst models, compared to $\sim30$--$40$ Myr in the no-burst case. At $R\sim1.8$--$2.3$ kpc, the quenching time increases to $\sim280\,(300)$ Myr for the $2(5)\%$ burst models, and $\sim100$ Myr without the burst. Our results support the outside-in quenching scenario inferred by \citet{Cramer_2019}, while probing larger galactocentric radii of $2.5-4.5$ kpc, where we find quenching times of $\sim 200 - 300$  Myr.

    \item RB 199, GMP 3016, and GMP 4232 are the least massive galaxies in the sample. Unlike the more massive systems, they are completely quenched in their discs. RPS can fully remove the gas reservoir of dwarf galaxies during their first pericentric passage \citep{Boselli_2008_dwarf_ellipticals}; these systems therefore represent galaxies in which the quenching process has acted rapidly and completely over $<300$ Myr. 
    The outermost region of GMP 3016 is not an exception. Due to the low SNR of the data, the $Q_\mathrm{AGE}$ posterior is very broad, and while the median value is $\approx 650$ Myr, values as low as 200 Myr are allowed at 1-sigma significance. We therefore conclude that the quenching age value at this distance from the galaxy center is nearly unconstrained. 
    

\end{itemize}

\paragraph{Galaxies in Norma}

\begin{itemize}
    
    \item ESO 137-001 is an intermediate-mass galaxy in the Norma cluster located at $r/r_{200}=0.18$. Its stripped tail extends for more than $\sim80$ kpc and has been detected in H$\alpha$, X-rays, and molecular gas \citep{Fossati_2016, Jachym_2019, Sun_2010}. 
    MUSE kinematic maps show that, while the stellar component remains dynamically undisturbed, the stripped gas preserves coherent rotational motions over several kpc, consistent with a purely hydrodynamical stripping interaction \citep{Fumagalli_2014}. From these kinematic signatures, these authors infer that the inner tail is fed by gas stripped $\approx 10-20$ Myr ago, in broad agreement with our estimates of the innermost quenched region of the disc that is very young ($Q_\mathrm{AGE}\approx 15-25$ Myr).
    
    \item ESO 137-002 is a massive spiral at $r/r_{200}=0.13$ from the Norma center. It presents clear signatures of edge-on ram-pressure stripping \citep{Zhang_2013_ESO002}, including a sharp H$\alpha$ truncation edge and an H$\alpha$ tail extending at least $\sim20$ kpc from the nucleus. X-ray observations reveal a $\sim40$ kpc tail characterized by unusually hot gas. Compared to ESO 137-001, ESO 137-002 is in a more advanced stage of the stripping process \citep{Sun_2010}. This is also confirmed by our results, showing that, at fixed galactocentric radius, ESO 137-002 has been stripped about 200 Myr before ESO 137-001. Moreover, ESO 137-002 exhibits a large value of $Q_\mathrm{AGE}/\tau_Q$, indicating a nearly instantaneous quenching at fixed radii.
\end{itemize}

\paragraph{Galaxies in A1367}

\begin{itemize}
    
    \item UGC 6697 (CGCG097-087) is located north-west of the A1367 cluster center at $r/r_{200}=0.38$. Its small velocity offset relative to the cluster indicates motion nearly in the plane of the sky, producing the remarkable edge-on ram-pressure stripping event with a $\sim100$ kpc ionized tail. 
    Despite the long tail of stripped gas, the galaxy disc is truncated only at large radii, suggesting that ram pressure has only recently begun to strongly affect the disc. The quenching times we derive are fully consistent with the other galaxies in the sample, once accounting for the position of the truncation radius. There are two factors that can contribute to the sudden onset of RPS: first the galaxy is currently traversing a shock front between two subcluster regions \citep{Ge_2019}, which could have increased the ICM density compared to a more relaxed cluster. Second, the galaxy has suffered a recent high-velocity interaction with the nearby companion CGCG097-087N \citep{Consolandi_2017}, which could also have contributed to an increased RPS efficiency from the perturbation of the gravitational potential.
    
    \item NGC 3860 (CGCG097-120) is a serendipitous projection of a cluster galaxy onto the Blue Infalling Group (BIG), from which it appears dynamically decoupled \citep{BlueInfallingGroup}. The galaxy is undergoing strong ram-pressure stripping during its infall into A1367 and it is very close, in projection, to the cluster center, at $r/r_{200}=0.08$. Its ionized gas disc is sharply truncated at $\sim6$ kpc, and the nearly azimuthally symmetric truncation suggests a face-on stripping geometry. Enhanced gas velocity dispersion and elevated BPT line ratios at the edge of the gaseous disc indicate stripping-induced shocks, whereas the inner disc still hosts active star formation with regular rotational kinematics \citep{BlueInfallingGroup}. 
    \citet{BlueInfallingGroup} studied the quenching times in a single region that is similar to the innermost region we studied. These authors found that quenching started $\sim570$ Myr ago and proceeded on a relatively gentle timescale of $\sim150$ Myr. At the larger radii probed here ($\sim 7-14$ kpc), we find older quenching ages of $\approx 2000$ Myr, consistent with an outside-in quenching pattern.
    This galaxy has the longest quenching ages of our sample. We explain this with a scenario in which it is currently after the most active stripping phase. Indeed it has no visible tail indicating a low stripping rate at present day. As a result, its stellar populations are aging throughout the whole stripped region. By comparing its $Q_\mathrm{AGE}$ profile with the average line shown in Figure \ref{fig:results_final}, we can estimate the bulk of the stripping occurred $\approx 700$ Myr ago. Lastly we note it is the only galaxy showing a decreasing trend of $\tau_Q$ with radius, suggestive of faster stripping in the outer (and older) regions compared to the inner ones.

\end{itemize}

\end{appendix}

\end{document}